\documentclass[11pt]{article}

\usepackage[]{acl}

\usepackage{times}
\usepackage{latexsym}

\usepackage[T1]{fontenc}

\usepackage[utf8]{inputenc}

\usepackage{microtype}

\usepackage{inconsolata}

\usepackage{graphicx}
\usepackage{amsmath}
\usepackage{amsfonts}
\usepackage{tcolorbox}
\tcbuselibrary{skins}
\usepackage{enumitem}
\usepackage{array}
\usepackage{booktabs}
\usepackage{multirow}
\usepackage{makecell}
\usepackage{subcaption}
\title{EviSnap: Faithful Evidence-Cited Explanations for Cold-Start Cross-Domain Recommendation}

\author{Yingjun Dai \\
  Carleton University\\
  ON, Canada\\
  \texttt{yingjundai@cmail.carleton.ca} \\\And
  Ahmed El-Roby \\
  Carleton University\\
  ON, Canada\\
  \texttt{ahmed.elroby@carleton.ca} \\}

\begin{document}
\maketitle
\begin{abstract}
Cold-start cross-domain recommender (CDR) systems predict a user’s preferences in a target domain using only their source-domain behavior, yet existing CDR models either map opaque embeddings or rely on post-hoc or LLM-generated rationales that are hard to audit. We introduce \textbf{EviSnap}, a lightweight CDR framework whose predictions are explained by construction with evidence-cited, faithful rationales. EviSnap distills noisy reviews into compact facet cards using an LLM offline, pairing each facet with verbatim supporting sentences. It then induces a shared, domain-agnostic concept bank by clustering facet embeddings and computes user-positive, user-negative, and item-presence concept activations via evidence-weighted pooling. A single linear concept-to-concept map transfers users across domains, and a linear scoring head yields per-concept additive contributions, enabling exact score decompositions and counterfactual 'what-if' edits grounded in the cited sentences. Experiments on the Amazon Reviews dataset across six transfers among Books, Movies, and Music show that EviSnap consistently outperforms strong mapping and review-text baselines while passing deletion- and sufficiency-based tests for explanation faithfulness.
\end{abstract}

\section{Introduction}
Real-world recommender systems frequently face cold-start users: people with interaction history in one domain (e.g., Movies) but none in a target domain (e.g., Music or Books). Cold-start cross-domain recommendation (CDR) tackles this by transferring preferences from a source domain to a target domain \citep{fernandez2012cross,khan2017cross,zang2022survey}. However, most CDR systems remain hard to audit. Mapping-based methods learn a transfer function between latent user embeddings and scale well \citep{man2017cross,hu2018conet,zhu2022personalized}, but the transferred signal is opaque: the model cannot clearly state what preferences were moved or why a target item is recommended. Review-aware models often improve accuracy by encoding text \citep{zheng2017joint,tay2018multi}, yet their explanations are typically post-hoc (e.g., attention/highlight rationales) and may not reflect the actual scoring function \citep{jain2019attention,wiegreffe2019attention,zhang2020explainable}. More recently, LLM-based recommenders can generate fluent justifications \citep{bao2023tallrec,mysore2023large,zhu2024collaborative}, but these can be costly at inference and are not guaranteed to be faithful or verifiable \citep{wu2024survey}.

We argue that CDR needs an intermediate representation that is (i) shared across domains, (ii) grounded in verifiable evidence, and (iii) used directly by the predictor so that explanations can be tested rather than narrated \citep{lei2016rationalizing,ross2017right,rudin2019stop}. To this end, we propose \textbf{EviSnap}, a lightweight CDR framework that produces faithful, evidence-cited explanations by construction. EviSnap operates in an evidence-grounded concept space built from reviews: an LLM distills raw reviews into compact facet cards (short, domain-agnostic facet phrases) and attaches verbatim supporting sentences. We embed facets from both domains and cluster them into a shared concept bank. For each user we compute separate positive and negative concept activations from praised vs.\ criticized evidence, and for each target item we compute concept presence activations from its evidence sentences. A single linear concept-to-concept map transfers users from the source concept space to the target space, and an additive linear scoring head produces ratings. Because the score is a sum of per-concept terms, EviSnap yields an exact score decomposition: the explanation is the model itself, paired with the highest-scoring user/item evidence sentences for each surfaced concept. The linear structure also enables transparent counterfactual edits (''what if this concept was stronger?'') with predictable changes in the score.

Our main contributions are:
\begin{itemize}[leftmargin=*, itemsep=2pt]
    \item \textbf{Evidence-cited, domain-agnostic concept representation for CDR.} We introduce an offline facet-card pipeline with verbatim evidence and induce a shared concept bank across domains, yielding sentence-traceable user-positive, user-negative, and item-presence activations.
    \item \textbf{Transparent transfer and faithful explanations by construction.} EviSnap uses a single linear concept mapping and an additive linear scorer, so reported per-concept contributions reconstruct the prediction exactly.
    \item \textbf{Empirical gains with faithfulness diagnostics.} Experiments over the Amazon Reviews dataset \citep{he2016ups} among \textsc{Books}, \textsc{Movies}, and \textsc{Music} , EviSnap outperforms strong mapping and text-based baselines while passing deletion- and sufficiency-based tests of explanation faithfulness \citep{lei2016rationalizing}.
\end{itemize}

\section{Problem Definition}
\label{sec:problem}

We study \emph{cross-domain recommendation (CDR)} for \emph{cold-start users} across two domains:
a source domain $S$ and a target domain $T$.
Let $\mathcal{U}$ be the user set and $\mathcal{I}_S$, $\mathcal{I}_T$ the item sets in $S$ and $T$.
We observe ratings $r_{ui}$ for some user--item pairs, and we use review text as side information:
$\mathcal{R}_S(u)$ denotes the set of source-domain reviews written by user $u$,
and $\mathcal{R}_T(i)$ denotes the set of target-domain reviews written about item $i$.

We adopt a \emph{user-level cold-start split}: users in training and test are disjoint.
At inference time for a test user $u$, we assume no target-domain history is available, i.e.,
only $\mathcal{R}_S(u)$ is observed for the user, while item-side text $\mathcal{R}_T(i)$ is available for target items.

Our goal is to learn a predictor $g_{\phi}$ that estimates a cold-start user's preference for a target item:
\[
\hat r_{ui} = g_{\phi}\!\left(u, i \mid \mathcal{R}_S(u), \mathcal{R}_T(i)\right),
\]
where $u \in \mathcal{U},\; i \in \mathcal{I}_T$.

\begin{figure*}[t]
  \centering
  \includegraphics[width=\textwidth]{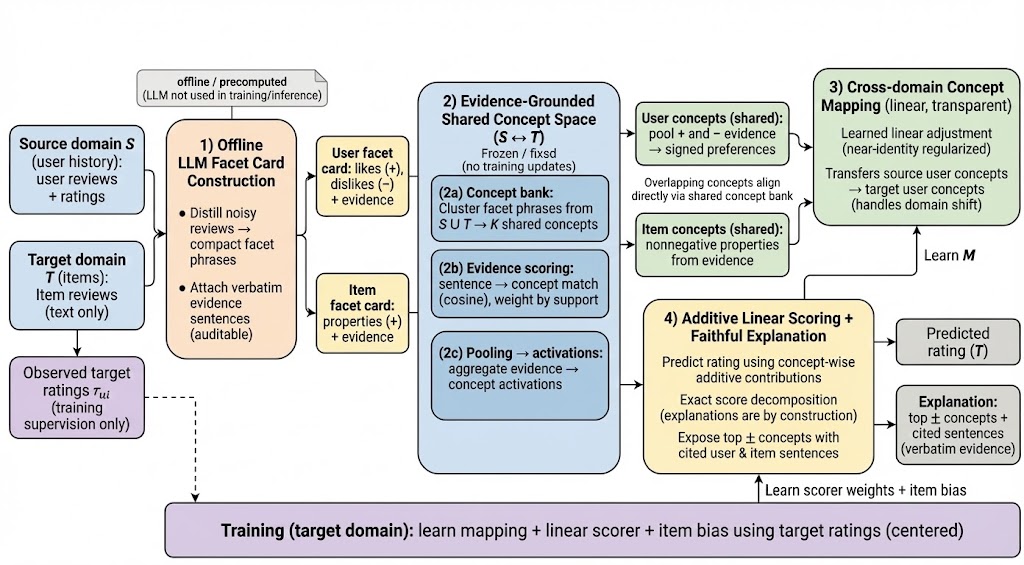}
  \vspace{-2mm}
  \caption{EviSnap overview: offline LLM facet cards with verbatim evidence, a frozen shared concept bank and activations, a near-identity linear transfer map $\mathbf{M}$, additive scoring with exact, evidence-cited per-concept contributions.}

  \vspace{-3mm}
  \label{fig:overview}
\end{figure*}

\section{Framework}
\label{sec:framework}

Given the cold-start CDR setting in Section~\ref{sec:problem}, we instantiate $g_{\phi}$ with an interpretable, evidence-grounded model that predicts target-domain ratings using a shared concept space as shown in Figure~\ref{fig:overview}.
EviSnap (i) distills reviews into evidence-cited facet cards, (ii) induces a shared concept bank and computes user/item concept activations, and (iii) applies a simple linear transfer and linear scoring head.
Because the final scorer is additive in concept features, the same quantities that produce $\hat r_{ui}$
also yield faithful, sentence-grounded explanations.

In this section, we describe each module and its training objective in turn.

\subsection{Generative Facet Card Construction with LLMs}
\label{sec:facet-cards}

\begin{figure*}[t]
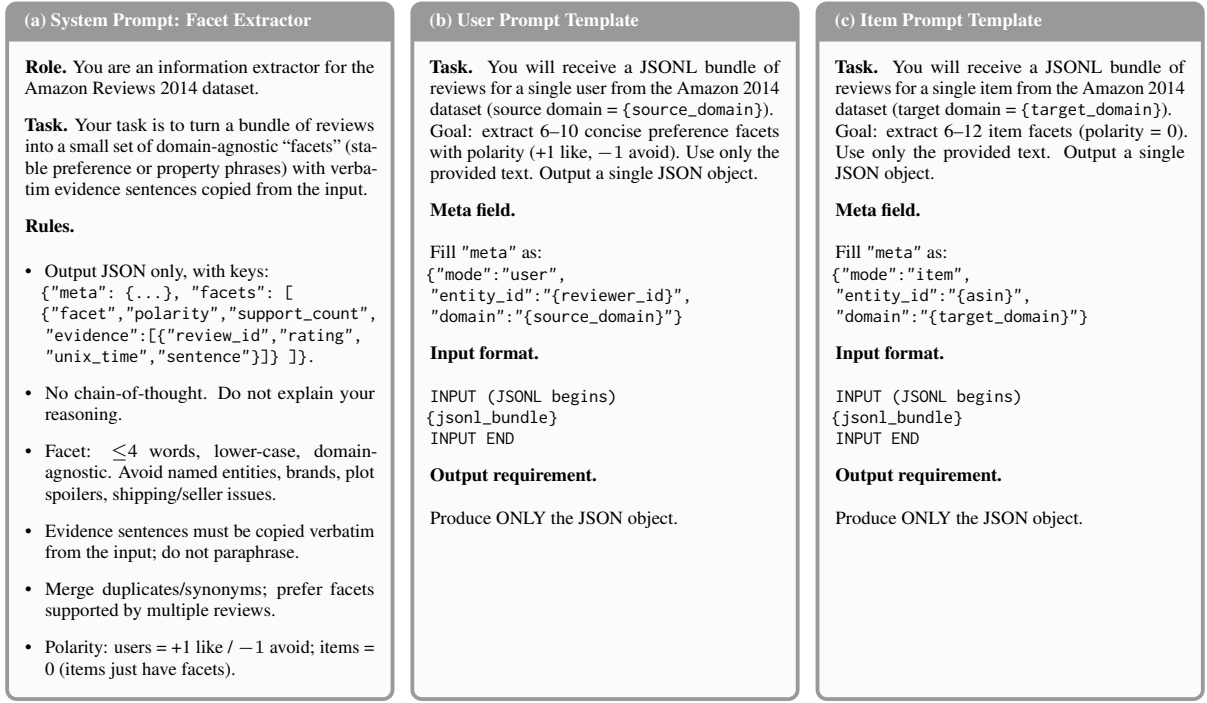

  \centering
  \scriptsize
  \setlength{\tabcolsep}{3pt}
  \begin{tabular}{@{}p{0.32\textwidth}p{0.32\textwidth}p{0.32\textwidth}@{}}

  \begin{tcolorbox}[
      colback=gray!3,
      colframe=black!40,
      title={(a) System Prompt: Facet Extractor},
      fonttitle=\bfseries,
      width=\linewidth,
      boxsep=3pt,
      left=3pt,right=3pt,top=4pt,bottom=4pt,
      equal height group=facetprompts
  ]
  \textbf{Role.} You are an information extractor for the Amazon Reviews 2014 dataset.

  \medskip
  \textbf{Task.} Your task is to turn a bundle of reviews into a small set of
  domain-agnostic ``facets'' (stable preference or property phrases) with
  verbatim evidence sentences copied from the input.

  \medskip
  \textbf{Rules.}
  \begin{itemize}[leftmargin=*, itemsep=1.5pt]
    \item Output JSON only, with keys:\\
    \texttt{\{"meta": \{...\}, "facets": [}\\
    \texttt{\{"facet","polarity","support\_count",}\\
    \texttt{"evidence":[\{"review\_id","rating", \\ "unix\_time","sentence"\}]\} ]\}.}
    \item No chain-of-thought. Do not explain your reasoning.
    \item Facet: $\leq$4 words, lower-case, domain-agnostic. Avoid named entities,
    brands, plot spoilers, shipping/seller issues.
    \item Evidence sentences must be copied verbatim from the input; do not paraphrase.
    \item Merge duplicates/synonyms; prefer facets supported by multiple reviews.
    \item Polarity: users = +1 like / $-1$ avoid; items = 0 (items just have facets).
  \end{itemize}
  \end{tcolorbox}
  &
  \begin{tcolorbox}[
      colback=gray!3,
      colframe=black!40,
      title={(b) User Prompt Template},
      fonttitle=\bfseries,
      width=\linewidth,
      boxsep=3pt,
      left=3pt,right=3pt,top=4pt,bottom=4pt,
      equal height group=facetprompts
  ]
  \textbf{Task.}
  You will receive a JSONL bundle of reviews for a single user from the Amazon 2014 dataset
  (source domain = \texttt{\{source\_domain\}}).\newline
  Goal: extract 6--10 concise preference facets with polarity (+1 like, $-1$ avoid).
  Use only the provided text. Output a single JSON object.

  \medskip
  \textbf{Meta field.}\\
  \\
  Fill \texttt{"meta"} as:\\
  {\ttfamily
  \noindent
  \{"mode":"user", "entity\_id":"\{reviewer\_id\}", "domain":"\{source\_domain\}"\}\par
  }

  \medskip
  \textbf{Input format.}\\
  \\
  {\ttfamily
  \noindent
  INPUT (JSONL begins)\\
  \{jsonl\_bundle\}\\
  INPUT END\par
  }

  \medskip
  \textbf{Output requirement.}\\
  \\
  Produce ONLY the JSON object.
  \end{tcolorbox}
  &
  \begin{tcolorbox}[
      colback=gray!3,
      colframe=black!40,
      title={(c) Item Prompt Template},
      fonttitle=\bfseries,
      width=\linewidth,
      boxsep=3pt,
      left=3pt,right=3pt,top=4pt,bottom=4pt,
      equal height group=facetprompts
  ]
  \textbf{Task.}
  You will receive a JSONL bundle of reviews for a single item from the Amazon 2014 dataset
  (target domain = \texttt{\{target\_domain\}}).\newline
  Goal: extract 6--12 item facets (polarity = 0). Use only the provided text.
  Output a single JSON object.

  \medskip
  \textbf{Meta field.}\\
  \\
  Fill \texttt{"meta"} as: \\
  {\ttfamily
  \noindent
  \{"mode":"item", \\ "entity\_id":"\{asin\}", "domain":"\{target\_domain\}"\}\par
  }

  \medskip
  \textbf{Input format.}\\
  \\
  {\ttfamily
  \noindent
  INPUT (JSONL begins)\\
  \{jsonl\_bundle\}\\
  INPUT END\par
  }

  \medskip
  \textbf{Output requirement.}\\
  \\
  Produce ONLY the JSON object.
  \end{tcolorbox}
  \\ 
  \end{tabular}

  \caption{LLM prompts used in our facet-extraction pipeline for the Amazon Reviews 2014 dataset:
  (a) system prompt defining the facet extraction role and JSON schema;
  (b) user prompt for user-level preference facets; and
  (c) item prompt for item-level facets.}
  \label{fig:facet-prompts-3}
\end{figure*}

We first preprocess review text into compact, auditable facet cards. For each source-domain user $u$ and each target-domain item $i$, we input the corresponding bundle of reviews to an LLM and obtain a single JSON object (Figure~\ref{fig:facet-prompts-3}). Each card contains a small set of short, domain-agnostic facet phrases paired with verbatim supporting sentences from the input reviews. Facets are accompanied by a support count.

User cards include facet polarity ($+1$ liked, $-1$ dislike), while item cards use polarity $0$ (items express properties only). We run this extraction offline once and treat facet phrases and evidence sentences as fixed inputs to EviSnap. The LLM is not used during model training or inference. This representation denoises long reviews while preserving traceability: every downstream concept activation can cite an original sentence.

\textbf{Example.}
User facets may include \emph{fast pacing} ($+1$) and \emph{slow plot} ($-1$) with copied evidence sentences; an item may include \emph{live energy} ($0$) with evidence such as ``The live drums give the songs amazing energy.''


\subsection{Evidence--Grounded Concept Space}

This stage builds a shared, interpretable feature space that supports both cross-domain transfer and faithful explanations.
We want the model’s internal variables to be (i) domain-agnostic so they align across $S$ and $T$,
(ii) evidence-grounded so each activation can be traced to specific sentences, and
(iii) simple enough that the final scoring function can decompose cleanly into per-concept contributions.
We achieve this by inducing a concept bank from facet phrases and computing user/item concept activations by pooling
sentence-to-concept evidence scores.

\paragraph{Concept bank.}
We induce concepts by clustering facet phrases so that synonymous facets collapse into a single dimension and the same concept labels can be reused across domains.
We embed every distinct facet phrase once using a frozen sentence encoder $f(\cdot)$:
\[
    \mathbf{e} = f(\text{facet phrase}) \in \mathbb{R}^d
\]
We then cluster all facet embeddings with $k$--means into $K$ clusters. The cluster centroids are our \emph{concept prototypes}
\[
    \mathbf{D} = 
    \begin{bmatrix}
        \mathbf{d}_1^\top \\
        \vdots\\
        \mathbf{d}_K^\top
    \end{bmatrix} \in \mathbb{R}^{K \times d}, 
    \quad \|\mathbf{d}_k\|_2 = 1 \;\text{for all }k
\]
Each concept $k$ is labeled by the facet phrase in its cluster whose embedding is closest to $\mathbf{d}_k$. These labels (e.g., ``\emph{live energy}'', ``\emph{vocal clarity}'') appear in our explanations.

\paragraph{Sentence--level evidence scores.}

We score concepts at the sentence level so that explanations can cite the single strongest supporting sentence (while still
accounting for multiple mentions). 

For any entity $e$ (user or item), we have a small set of evidence sentences
\[
\mathcal{S}(e) = \{(s_j, w_j)\}_{j=1}^{N_e},
\]
where $s_j$ is an evidence sentence and $w_j$ is a non--negative weight. We use
\[
    w_j = \log\bigl(1 + \text{support count}(s_j)\bigr)
\]
so that facets mentioned in more reviews carry slightly more weight.

We embed each sentence,
\[
\mathbf{h}_j = f(s_j) \in \mathbb{R}^d, \qquad \|\mathbf{h}_j\|_2 = 1,
\]
and compute its alignment with each concept prototype $\mathbf{d}_k$ using cosine similarity followed by ReLU:
\[
\alpha_{jk} = \max\left(0, \;\mathbf{h}_j^\top \mathbf{d}_k\right)
\]
Here $\alpha_{jk}\in[0,1]$ measures how strongly sentence $s_j$ supports concept $k$. Negative or unrelated evidence is clipped to $0$.

\paragraph{Pooling over evidence.}
This pooling choice lets us produce explanations by pointing to the highest-scoring evidence sentence for each concept,
while still rewarding concepts supported by multiple reviews.

To obtain a fixed-length concept vector from a variable number of evidence sentences, we aggregate sentence-level alignments into one score per concept.
We use weighted log--sum--exp pooling as a smooth max that highlights the strongest supporting sentence without discarding additional supporting evidence:
\[
s_k(e) = \frac{1}{\alpha} \log \sum_j \tilde{w}_j \,\exp\!\big(\alpha\, \alpha_{jk}\big),
\quad \tilde{w}_j = \frac{w_j}{\sum_j w_j},
\]
where $\alpha>0$ is a temperature. When $\alpha$ is large, this behaves like a soft maximum over sentences.

\paragraph{User and item concept vectors.}

We represent preference polarity explicitly for users (likes vs.\ dislikes) because it affects whether a concept should increase or decrease a target-item score, whereas items only express presence of properties.

For users, facet cards differentiate positive and negative opinions. We pool them separately:
\begin{itemize}
    \item $\mathbf{U}^+(u) \in \mathbb{R}^K$ from positive evidence sentences (facets the user likes),
    \item $\mathbf{U}^-(u) \in \mathbb{R}^K$ from negative evidence sentences (facets the user dislikes).
\end{itemize}
We then form a \emph{signed} source--domain user vector:
\[
\mathbf{a}_S(u) = \mathbf{U}^+(u) - \mathbf{U}^-(u)
\]
Each entry is high and positive if the user repeatedly praises that concept, and negative if they repeatedly criticize it.

For items in the target domain, facets describe item attributes rather than preferences.
Therefore item facets have no polarity (we set polarity to 0) and we pool all item evidence sentences into a nonnegative concept vector:
\[
\mathbf{b}(i) \in \mathbb{R}^K_{\ge 0},
\]
which encodes which concepts the item offers and how strongly.

\subsection{Cross--Domain Concept Mapping}

Cold-start users have no history in the target domain, so we must transfer their preferences from $S$ into a target-domain
concept representation that can be compared with target items.
We choose a single linear map for transparency: its weights directly show which source concepts contribute to which
target concepts, making the transfer step itself inspectable rather than a black-box embedding shift.

We now map each user's source domain concept vector to a target domain vector. We adopt a simple linear map
\[
\mathbf{a}_T(u) = \mathbf{M}\,\mathbf{a}_S(u), \qquad \mathbf{M} \in \mathbb{R}^{K \times K}
\]
Matrix $\mathbf{M}$ is learned from data and initialized to the identity. Intuitively, $\mathbf{M}$ says how source concepts (e.g., ``\emph{fast--paced action}'') translate into target concepts (e.g., ``\emph{live energy}'').

To keep $\mathbf{M}$ easy to interpret and to avoid overfitting in the cold--start setting, we regularize it toward the identity:
\[
\|\mathbf{M} - \mathbf{I}\|_F^2,
\]
where $\mathbf{I}$ is the $K\times K$ identity matrix and $\|\cdot\|_F$ denotes the Frobenius norm.

\subsection{Linear Rating Prediction}

Given mapped user concepts $\mathbf{a}_T(u)$ and item concepts $\mathbf{b}(i)$, we need a scoring function that is
(i) accurate enough to model user-item matching, but also (ii) additively decomposable so explanations can be faithful.
We therefore use a linear head over per-concept features: an interaction term captures match (user preference aligned with item property), while marginal terms capture user and item specific tendencies. This keeps inference lightweight and ensures each concept’s effect on the score can be computed exactly. 

We construct a feature vector using an element wise interaction and optional marginal terms:

\begin{align*}
\mathbf{x}^{(\text{int})}(u,i) &= \mathbf{a}_T(u) \odot \mathbf{b}(i) \;\;\in\mathbb{R}^K,\\
\mathbf{x}^{(u)}(u,i) &= \mathbf{a}_T(u) \in\mathbb{R}^K,\\
\mathbf{x}^{(i)}(u,i) &= \mathbf{b}(i) \in\mathbb{R}^K,
\end{align*}
where $\odot$ denotes element--wise multiplication. We then concatenate them:
\[
\mathbf{z}(u,i) = \big[\;\mathbf{x}^{(\text{int})}(u,i) \;\big|\; \mathbf{x}^{(u)}(u,i) \;\big|\; \mathbf{x}^{(i)}(u,i)\;\big],
\]
where $\mathbf{z}(u,i) \in \mathbb{R}^{3K}$.

A single linear layer computes a \emph{centered} rating score:
\[
y_c(u,i) = \mathbf{w}^\top \mathbf{z}(u,i) + b_i,
\]
where $\mathbf{w} \in \mathbb{R}^{3K}$ are the head weights and $b_i \in \mathbb{R}$ is an item bias parameter (one scalar per item). The final predicted rating adds back the target--domain mean rating $\mu_T$:
\[
\hat r_{ui} = \mu_T + y_c(u,i)
\]

In practice, we compute the mean rating $\mu_T$ over the training portion of the target domain and train on \emph{centered} labels $r_{ui} - \mu_T$. At evaluation time, we clamp $\hat r_{ui}$ to the valid rating range (e.g., $[1,5]$).

\paragraph{Per--concept contributions (for explanation).}
The head weights $\mathbf{w}$ can be seen as three blocks:
\[
\mathbf{w} = \big[\;\mathbf{w}^{(\text{int})}\;\big|\;\mathbf{w}^{(u)}\;\big|\;\mathbf{w}^{(i)}\;\big],
\]
each of length $K$. The centered score can then be written as a sum over concepts:

\[
\begin{aligned}
y_c(u,i)
&= \sum_{k=1}^K \Big(
    w^{(\text{int})}_k\,a_{T,k}(u)\,b_k(i)
    + w^{(u)}_k\,a_{T,k}(u) \\
&\qquad\qquad\quad
    +\, w^{(i)}_k\,b_k(i)
\Big) + b_i
\end{aligned}
\]

For explanation, we define the per--concept contribution
\[
\begin{aligned}
\text{contrib}_k(u,i) 
= w^{(\text{int})}_k\,a_{T,k}(u)\,b_k(i) \\ 
+ \ w^{(u)}_k\,a_{T,k}(u) 
+ w^{(i)}_k\,b_k(i),
\end{aligned}
\]
and we display the top few positive and negative $\text{contrib}_k$ together with the concept label and the best--matching user and item evidence sentences. 
Crucially, the explanation is \emph{faithful by construction}: there is no separate explanation module---$\text{contrib}_k(u,i)$ is defined to be exactly the $k$-th additive term in the model's scoring function.
Therefore the centered score decomposes exactly as
\[
y_c(u,i) = \sum_k \text{contrib}_k(u,i) + b_i,
\]
so the reported contributions (plus the item bias $b_i$ and the mean $\mu_T$) reconstruct the predicted rating exactly.

\subsection{Training }

We train only the transfer and scoring parameters: the cross-domain map $\mathbf{M}$, the linear head weights $\mathbf{w}$,
and item biases $b_i$, so that predicted target-domain ratings fit observed ratings while keeping transfer interpretable and stable.
We center ratings to factor out the global target-domain mean, and we apply light regularization to (i) keep $\mathbf{M}$ near-identity
to reduce overfitting and preserve concept semantics, and (ii) prevent item biases from dominating the explanation.

Let $\mathcal{D}_{\text{train}} \subset \mathcal{U} \times \mathcal{I}_T$ denote the training set of user--item pairs in the target domain, with observed ratings $r_{ui}$. We use a user--level cold--start split: users in training and test do not overlap.

We first compute the mean target--domain rating over training data,
\[
\mu_T = \frac{1}{|\mathcal{D}_{\text{train}}|}\sum_{(u,i)\in\mathcal{D}_{\text{train}}} r_{ui}
\]
We then train the model to predict the \emph{centered} rating $r_{ui} - \mu_T$. The main loss term is mean squared error:
\[
\mathcal{L}_{\text{MSE}} 
= \frac{1}{|\mathcal{D}_{\text{train}}|}
\sum_{(u,i)\in\mathcal{D}_{\text{train}}}
\big(y_c(u,i) - (r_{ui} - \mu_T)\big)^2
\]

We add light regularization to stabilize the mapping and keep biases small:
\[
\mathcal{L}_{\text{reg}} 
= \lambda_M \|\mathbf{M} - \mathbf{I}\|_F^2
+ \lambda_b \sum_i b_i^2,
\]
where $\lambda_M$ and $\lambda_b$ are small hyperparameters.

The final training objective is
\[
\mathcal{L} = \mathcal{L}_{\text{MSE}} + \mathcal{L}_{\text{reg}}
\]

We optimize $\mathcal{L}$ with mini--batch stochastic gradient descent (AdamW in our implementation). Because all operations are linear in the learnable parameters (except for the fixed encoder and concept scoring), training is stable and efficient.

\section{Experimental Evaluation}
\label{sec:experiments}

We evaluate EviSnap in the cold-start cross-domain recommendation setting on review-driven transfers between \textsc{Movies}, \textsc{Books}, and \textsc{Music}. This section describes the dataset construction and evaluation scenarios, and then introduces the baselines used for comparison.

\begin{table*}[h]
\centering
\caption{Cross-domain recommendation results.
Best and second best are in \textbf{bold} and \underline{underlined}, respectively.}
\label{tab:crossdomain_results}
\resizebox{\linewidth}{!}{%
\begin{tabular}{llcccccc}
\toprule
\textbf{Method} & \textbf{Metric} &
\makecell{\textbf{Books}\\\textbf{$\rightarrow$ Music}} &
\makecell{\textbf{Books}\\\textbf{$\rightarrow$ Movies}} &
\makecell{\textbf{Movies}\\\textbf{$\rightarrow$ Music}} &
\makecell{\textbf{Movies}\\\textbf{$\rightarrow$ Books}} &
\makecell{\textbf{Music}\\\textbf{$\rightarrow$ Movies}} &
\makecell{\textbf{Music}\\\textbf{$\rightarrow$ Books}} \\
\midrule
\multirow[c]{2}{*}{EMCDR}
 & MAE  & 1.2894 & 0.9701 & 1.1073 & 0.9834 & 0.9860 & 1.1730 \\
 & RMSE & 1.5811 & 1.2461 & 1.3679 & 1.2427 & 1.2856 & 1.4954 \\
\midrule
\multirow[c]{2}{*}{PTUPCDR}
 & MAE  & 1.0473 & 0.9453 & 0.9384 & 0.9278 & 0.9642 & 1.0809 \\
 & RMSE & 1.3794 & 1.2326 & 1.2562 & 1.2073 & 1.2760 & 1.4276 \\
\midrule
\multirow[c]{2}{*}{MACDR}
 & MAE  & 0.8397 & 0.8987 & 0.8757 & 0.8790 & 0.9038 & 0.8579 \\
 & RMSE & 1.1042 & 1.1388 & 1.1356 & 1.1376 & 1.1564 & 1.0921 \\
\midrule
\multirow[c]{2}{*}{HeroGraph}
 & MAE  & 0.8150 & 0.8610 & \underline{0.7980} & 0.8670 & \textbf{0.8020} & 0.8860 \\
 & RMSE & \underline{1.0260} & 1.1180 & 1.1010 & 1.1330 & \textbf{1.0880} & 1.1210 \\
\midrule
\multirow[c]{2}{*}{DeepCoNN$+$}
 & MAE  & \underline{0.8064} & \underline{0.8462} & 0.8413 & \underline{0.7980} & 0.9254 & \underline{0.8548} \\
 & RMSE & 1.0514 & \underline{1.0919} & \underline{1.0953} & \underline{1.0180} & 1.1777 & \underline{1.0736} \\
\midrule 
\multirow[c]{2}{*}{\textbf{EviSnap}}
 & MAE  & \textbf{0.7882} & \textbf{0.8205} & \textbf{0.7768} & \textbf{0.7916} & \underline{0.8990} & \textbf{0.8298} \\
 & RMSE & \textbf{1.0243} & \textbf{1.0696} & \textbf{1.0438} & \textbf{1.0056} & \underline{1.1427} & \textbf{1.0446} \\
\bottomrule
\end{tabular}}
\end{table*}

\subsection{Dataset and Transfer Scenarios}
\label{sec:dataset}

We use the Amazon Reviews 2014 dataset~\citep{he2016ups} and consider the \textsc{Books}, \textsc{Movies}, and \textsc{Music} domains. Following common practice in cross-domain recommendation, we evaluate all six directed transfers $S\!\rightarrow\!T$ among these domains.

For each $S\!\rightarrow\!T$, we restrict to overlapping users who have source-domain text $\mathcal{R}_S(u)$ and at least one observed rating on a target-domain item. We then create a user-level cold-start split by randomly assigning 80\% of these users to training and 20\% to testing (users are disjoint across splits). At test time, the model observes only $\mathcal{R}_S(u)$ for the user (no target-domain user history), while target-item text $\mathcal{R}_T(i)$ is available for $i\in\mathcal{I}_T$; test ratings are used only for evaluation.

To prevent leakage through item-side text, when constructing $\mathcal{R}_T(i)$ (and extracting target-item facet cards) we exclude all target-domain reviews authored by held-out users, so the model never observes target-domain review text from test users.

\subsection{Experimental Setting}
\label{sec:exp_setting}

We use precomputed facet cards for source-domain users and target-domain items, constructed per $S\!\rightarrow\!T$ task using the split and leakage prevention protocol in Section~\ref{sec:dataset}, and treat them as fixed inputs (no LLM calls during model training/evaluation). Facet phrases and evidence sentences are embedded with a frozen sentence encoder, clustered with $k$-means to form a shared $K$-concept bank, and pooled into user/item concept activations via evidence-weighted log-sum-exp pooling over ReLU cosine similarities. We train only the linear concept map $\mathbf{M}$ and the additive linear head by minimizing MSE on centered target-domain ratings, regularizing $\mathbf{M}$ toward identity and applying a small $\ell_2$ penalty on item biases. We use a single hyperparameter setting for all transfers and do not perform hyperparameter tuning. We ran each scenario 5 times and took the average.

\subsection{Baseline Methods}
\label{sec:baselines}

We compare against five cross-domain recommendation baselines that cover both mapping-based transfer and review-text models:

\begin{itemize}[leftmargin=*,itemsep=3pt]
    \item \textbf{EMCDR}~\citep{man2017cross}: learns latent representations in each domain and trains a mapping function to transfer user representations from $S$ to $T$.
    \item \textbf{PTUPCDR}~\citep{zhu2022personalized}: It uses a meta-network to generate personalized transfer (bridge) functions for different users.
    \item \textbf{MACDR}~\citep{wang2024making}: It employs a prototype-enhanced mixture-of-experts transfer mechanism and distribution alignment to improve robustness under sparse supervision.
    \item \textbf{DeepCoNN+}~\citep{zheng2017joint}: It is a text-based recommender that models users and items from review text; we use its cross-domain variant with an added mapping layer for transfer.
    \item \textbf{HeroGraph}~\cite{cui2020herograph}: It obtains cross-domain information by a shared graph, which is constructed by collecting users' and items' information from multiple domains.
\end{itemize}

\begin{figure*}[t]
  \centering
  \includegraphics[width=\textwidth]{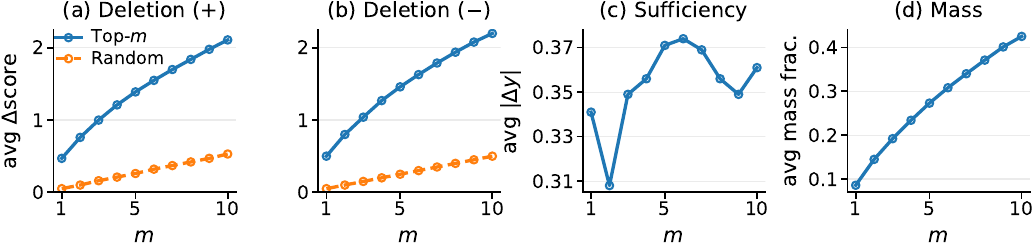}
  \vspace{-2mm}
  \caption{Faithfulness diagnostics on \textsc{Music}$\rightarrow$\textsc{Movies}.
  (a) positive deletion vs.\ random, (b) negative deletion vs.\ random,
  (c) sufficiency ($|y_{\text{full}}-y_m|$), (d) contribution mass.}
  \label{fig:faith_1x4_music_movies}
  \vspace{-4mm}
\end{figure*}


\subsection{Main Results: Cold-Start Cross-Domain Recommendation}
\label{sec:main-results}

Table~\ref{tab:crossdomain_results} reports MAE/RMSE on six directed transfers among
\textsc{Books}, \textsc{Movies}, and \textsc{Music} under the user-level cold-start split.
EviSnap achieves the best performance on five of six transfer directions, and is second-best on the remaining \textsc{Music}$\rightarrow$\textsc{Movies} setting, where HeroGraph is strongest (MAE $0.8020$, RMSE $1.0880$).
Averaged over transfers, EviSnap improves over the strongest review-text baseline DeepCoNN+ from $0.845$ to $0.818$ MAE and from $1.085$ to $1.055$ RMSE (relative $3.3\%$ and $2.7\%$), and yields larger gains over the best mapping-based baseline MACDR ($6.6\%$ MAE, $6.4\%$ RMSE).
The largest improvements occur on \textsc{Movies}$\rightarrow$\textsc{Music}, where EviSnap reduces MAE from $0.8413$ to $0.7768$ and RMSE from $1.0953$ to $1.0438$.
Overall, these results suggest that transferring users through an evidence-grounded concept space is competitive for cold-start CDR while maintaining an additive structure that directly supports faithful, sentence-grounded explanations.

\subsection{Ablation: Linear-Head Feature Blocks}
\label{sec:ablation_head}

Our scorer is designed to be both accurate and decomposable into per-concept effects. To isolate whether gains come primarily from \emph{concept-level matching} or from \emph{marginal} user/item signals,
we ablate which concept-feature blocks are fed into the linear head, while keeping the concept bank, pooling, mapping $\mathbf{M}$, item bias, and training objective fixed.

Given mapped user concepts $\mathbf{a}_T(u)\in\mathbb{R}^K$ and item concepts $\mathbf{b}(i)\in\mathbb{R}^K_{\ge 0}$,
we define three blocks:
\[
\label{eq:head_blocks}
\mathbf{x}^{(\mathrm{int})}(u,i)=\mathbf{a}_T(u)\odot\mathbf{b}(i),
\]
\[
\mathbf{x}^{(u)}(u,i)=\mathbf{a}_T(u),
\]
\[
\mathbf{x}^{(i)}(u,i)=\mathbf{b}(i)
\]
We then form the head input using binary switches $\delta_u,\delta_i\in\{0,1\}$:
\[
\label{eq:z_switch}
\mathbf{z}(u,i)=\Big[\ \mathbf{x}^{(\mathrm{int})}(u,i)\ \Big|\ \delta_u\,\mathbf{x}^{(u)}(u,i)\ \Big|\ \delta_i\,\mathbf{x}^{(i)}(u,i)\ \Big]
\]
This yields four variants: \textbf{IntOnly} $(\delta_u{=}0,\delta_i{=}0)$, \textbf{Int+User} $(1,0)$,
\textbf{Int+Item} $(0,1)$, and \textbf{Full} $(1,1)$.

\vspace{-1mm}
\begin{table}[t]
\centering
\small
\setlength{\tabcolsep}{6pt}
\begin{tabular}{@{}lccc cc@{}}
\toprule
\multirow{2}{*}{\textbf{Variant}} & \multicolumn{3}{c}{\textbf{Blocks in $\mathbf{z}(u,i)$}} & \multirow{2}{*}{\textbf{MAE}$\downarrow$} & \multirow{2}{*}{\textbf{RMSE}$\downarrow$} \\
\cmidrule(lr){2-4}
 & $\mathbf{x}^{(\mathrm{int})}$ & $\mathbf{x}^{(u)}$ & $\mathbf{x}^{(i)}$ &  &  \\
\midrule
IntOnly  & 1 & 0 & 0 & 0.9170 & 1.1522 \\
Int+User & 1 & 1 & 0 & 0.9155 & 1.1509 \\
Int+Item & 1 & 0 & 1 & \underline{0.9079} & \underline{1.1504} \\
Full     & 1 & 1 & 1 & \textbf{0.9013} & \textbf{1.1468} \\
\bottomrule
\end{tabular}
\vspace{-1mm}
\caption{Linear-head block ablation on Music to Movies ($K{=}128$). Best/second-best are in \textbf{bold}/\underline{underline}.}
\label{tab:head_ablation_music_movies}
\vspace{-2mm}
\end{table}

On \textsc{Music} to \textsc{Movies} (other transfer directions show similar trends), \textbf{Full} performs best, reducing error relative to \textbf{IntOnly} by $0.0157$ MAE and $0.0054$ RMSE, indicating that marginal blocks provide additional signal beyond pure element-wise matching.




\subsection{Faithfulness Diagnostics}
\label{sec:faithfulness}

Because $y_c(u,i)$ is additive in concept terms (Section~\ref{sec:framework}), we can test faithfulness via direct concept-space interventions. On \textsc{Music} to \textsc{Movies} (other transfers have similar behaviors), we report (i) \emph{deletion}: ablate the top-$m$ positive/negative concepts ranked by $\text{contrib}_k$ and measure the resulting score change, using random deletions as a control; and (ii) \emph{sufficiency}: keep only the top-$m$ concepts by $|\text{contrib}_k|$ and compute the residual $|y_c-y_c^{(m)}|$. Figure~\ref{fig:faith_1x4_music_movies} shows that top-$m$ deletion perturbs predictions far more than random, while a small set of top concepts largely reconstructs $y_c$, supporting that the surfaced concepts are the model’s true drivers and that decisions are distributed across multiple facets.

\subsection{Qualitative Analysis}
\label{sec:qualitative}


Table~\ref{tab:qual_case_movies_music} illustrates a cold-start \textsc{Movies} to \textsc{Music} explanation for user \texttt{ALLHLOG4NLA0A} and item \texttt{B0007SMCWY}. Each row corresponds to a concept $k$ in the shared concept bank. The reported score is the exact additive term $\text{contrib}_k(u,i)$ in our linear head, so positive (negative) values raise (lower) the predicted rating by that amount. For each concept, we cite the highest-alignment verbatim sentence from the user's \textsc{Movies} reviews and the item’s \textsc{Music} reviews, making the transfer auditable at the sentence level. In this case, the prediction is supported by aligned evidence for \emph{musicianship/deep cuts} (the user values technical skill. the album provides many lesser-known performances), as well as complementary signals of \emph{great value} and \emph{nostalgia}. For readability, we show only the largest-magnitude contributions.

\vspace{-2mm}
\begin{table}[t]
\centering
\scriptsize
\setlength{\tabcolsep}{3pt}
\renewcommand{\arraystretch}{1.12}
\caption{One \textsc{Movies}$\rightarrow$\textsc{Music} explanation}
\label{tab:qual_case_movies_music}
\begin{tabular}{@{}p{0.3\linewidth}p{0.66\linewidth}@{}}
\toprule
\textbf{Concept (contrib)} & \textbf{Cited evidence (user $\rightarrow$ item)} \\
\midrule
\textbf{musicianship/deep cuts} \newline
(+0.45) &
\textit{User:} \detokenize{A MUST FOR EVERY MUSICIAN!}\newline
\textit{Item:} \detokenize{The 38 tracks are non-sequential and they include many lesser-known performances -- both originals and covers -- in addition to most of the well-known hits,}
\\
\textbf{great value} \newline
(+0.32) &
\textit{User:} \detokenize{This is worth every singe penny.}\newline
\textit{Item:} \detokenize{But the price can't be beat: the set is currently available new from Amazon Marketplace vendors for around \$5 (plus shipping).}
\\
\textbf{nostalgia} \newline
(+0.28) &
\textit{User:} \detokenize{It brings me back to the 80s when you bought an Lp and the WHOLE LP was very good.}\newline
\textit{Item:} \detokenize{I am a hugh Al Green fan and this set brings back a lot of memories and still makes me want to dance and grove.}
\\
\bottomrule
\end{tabular}
\vspace{-3mm}
\end{table}

\section{Related Work}
\label{sec:related-work}

Cold-start cross-domain recommendation (CDR) transfers a user’s signal from a source to a target domain, typically by learning domain-specific representations and a mapping/bridge function (e.g., EMCDR/CoNet and later personalized or prototype/mixture-based transfer) to cope with sparse supervision \citep{zang2022survey,man2017cross,hu2018conet,zhu2022personalized,wang2024making}. Reviews provide fine-grained signals via aspect-style models and neural review encoders \citep{zheng2017joint,tay2018multi}, but many “explanations” in review-aware recommenders are post-hoc (e.g., attention/highlights) and need not reflect the true scoring rule \citep{jain2019attention,wiegreffe2019attention}; similarly, LLM-generated justifications can be fluent yet weakly coupled to the predictor and hard to audit \citep{bao2023tallrec,wu2024survey}. EviSnap follows interpretability-by-construction \citep{rudin2019stop}: it distills reviews into facet cards with verbatim evidence and predicts in an additive concept space with a linear transfer, yielding exact per-concept contributions that can be directly validated via deletion/sufficiency interventions \citep{lei2016rationalizing,ross2017right}.

\section{Conclusion}

We propose EviSnap, a lightweight cold-start cross-domain recommender that transfers users through an evidence-grounded concept space derived from reviews. An offline LLM distills reviews into facet cards with verbatim evidence, and a shared concept bank with a linear map and additive head yields ratings with exact, evidence-cited per-concept contributions. Across six Amazon Reviews transfers among \textsc{Books}, \textsc{Movies}, and \textsc{Music}, EviSnap outperforms SOTA baselines, and shows the faithfulness of the surfaced concepts transfers.

\section{Limitations}
EviSnap relies on an offline LLM step to distill reviews into facet cards. While evidence sentences are verbatim, the extracted facet phrases and user polarity labels may be sensitive to the specific LLM and prompts and may inherit noise or biases from the model and data. Our approach also assumes sufficient review text for both source-domain users and target-domain items. Performance and explanation quality may degrade in text-sparse settings where users/items have few or no reviews. Finally, interpretability is enabled by an unsupervised $k$-means concept bank and a linear mapping/additive head. Concept quality can depend on the encoder and the choice of $K$, and the linear form may miss higher-order interactions that could benefit some transfers.

\bibliography{custom}

@inproceedings{man2017cross,
  title={Cross-domain recommendation: An embedding and mapping approach.},
  author={Man, Tong and Shen, Huawei and Jin, Xiaolong and Cheng, Xueqi},
  booktitle={IJCAI},
  volume={17},
  pages={2464--2470},
  year={2017}
}

@inproceedings{hu2018conet,
  title={Conet: Collaborative cross networks for cross-domain recommendation},
  author={Hu, Guangneng and Zhang, Yu and Yang, Qiang},
  booktitle={Proceedings of the 27th ACM international conference on information and knowledge management},
  pages={667--676},
  year={2018}
}

@inproceedings{zhu2022personalized,
  title={Personalized transfer of user preferences for cross-domain recommendation},
  author={Zhu, Yongchun and Tang, Zhenwei and Liu, Yudan and Zhuang, Fuzhen and Xie, Ruobing and Zhang, Xu and Lin, Leyu and He, Qing},
  booktitle={Proceedings of the fifteenth ACM international conference on web search and data mining},
  pages={1507--1515},
  year={2022}
}

@inproceedings{zheng2017joint,
  title={Joint deep modeling of users and items using reviews for recommendation},
  author={Zheng, Lei and Noroozi, Vahid and Yu, Philip S},
  booktitle={Proceedings of the tenth ACM international conference on web search and data mining},
  pages={425--434},
  year={2017}
}

@article{jain2019attention,
  title={Attention is not explanation},
  author={Jain, Sarthak and Wallace, Byron C},
  journal={arXiv preprint arXiv:1902.10186},
  year={2019}
}

@inproceedings{fernandez2012cross,
  title={Cross-domain recommender systems: A survey of the state of the art},
  author={Fern{\'a}ndez-Tob{\'\i}as, Ignacio and Cantador, Iv{\'a}n and Kaminskas, Marius and Ricci, Francesco},
  booktitle={Spanish conference on information retrieval},
  volume={24},
  year={2012},
  organization={sn}
}

@article{khan2017cross,
  title={Cross domain recommender systems: A systematic literature review},
  author={Khan, Muhammad Murad and Ibrahim, Roliana and Ghani, Imran},
  journal={ACM Computing Surveys (CSUR)},
  volume={50},
  number={3},
  pages={1--34},
  year={2017},
  publisher={ACM New York, NY, USA}
}

@article{zang2022survey,
  title={A survey on cross-domain recommendation: taxonomies, methods, and future directions},
  author={Zang, Tianzi and Zhu, Yanmin and Liu, Haobing and Zhang, Ruohan and Yu, Jiadi},
  journal={ACM Transactions on Information Systems},
  volume={41},
  number={2},
  pages={1--39},
  year={2022},
  publisher={ACM New York, NY}
}

@inproceedings{tay2018multi,
  title={Multi-pointer co-attention networks for recommendation},
  author={Tay, Yi and Luu, Anh Tuan and Hui, Siu Cheung},
  booktitle={Proceedings of the 24th ACM SIGKDD international conference on knowledge discovery \& data mining},
  pages={2309--2318},
  year={2018}
}

@article{zhang2020explainable,
  title={Explainable recommendation: A survey and new perspectives},
  author={Zhang, Yongfeng and Chen, Xu and others},
  journal={Foundations and Trends{\textregistered} in Information Retrieval},
  volume={14},
  number={1},
  pages={1--101},
  year={2020},
  publisher={Now Publishers, Inc.}
}

@inproceedings{wiegreffe2019attention,
  title={Attention is not not Explanation},
  author={Wiegreffe, Sarah and Pinter, Yuval},
  booktitle={Proceedings of the 2019 Conference on Empirical Methods in Natural Language Processing and the 9th International Joint Conference on Natural Language Processing (EMNLP-IJCNLP)},
  pages={11--20},
  year={2019}
}

@inproceedings{lei2016rationalizing,
  title={Rationalizing Neural Predictions},
  author={Lei, Tao and Barzilay, Regina and Jaakkola, Tommi},
  booktitle={Proceedings of the 2016 Conference on Empirical Methods in Natural Language Processing},
  pages={107--117},
  year={2016}
}

@inproceedings{ross2017right,
  title={Right for the Right Reasons: Training Differentiable Models by Constraining their Explanations},
  author={Ross, Andrew Slavin and Hughes, Michael C and Doshi-Velez, Finale},
  booktitle={IJCAI},
  year={2017}
}

@article{wu2024survey,
  title={A survey on large language models for recommendation},
  author={Wu, Likang and Zheng, Zhi and Qiu, Zhaopeng and Wang, Hao and Gu, Hongchao and Shen, Tingjia and Qin, Chuan and Zhu, Chen and Zhu, Hengshu and Liu, Qi and others},
  journal={World Wide Web},
  volume={27},
  number={5},
  pages={60},
  year={2024},
  publisher={Springer}
}

@inproceedings{bao2023tallrec,
  title={Tallrec: An effective and efficient tuning framework to align large language model with recommendation},
  author={Bao, Keqin and Zhang, Jizhi and Zhang, Yang and Wang, Wenjie and Feng, Fuli and He, Xiangnan},
  booktitle={Proceedings of the 17th ACM conference on recommender systems},
  pages={1007--1014},
  year={2023}
}

@inproceedings{mysore2023large,
  title={Large language model augmented narrative driven recommendations},
  author={Mysore, Sheshera and McCallum, Andrew and Zamani, Hamed},
  booktitle={Proceedings of the 17th ACM conference on recommender systems},
  pages={777--783},
  year={2023}
}

@inproceedings{zhu2024collaborative,
  title={Collaborative large language model for recommender systems},
  author={Zhu, Yaochen and Wu, Liang and Guo, Qi and Hong, Liangjie and Li, Jundong},
  booktitle={Proceedings of the ACM Web Conference 2024},
  pages={3162--3172},
  year={2024}
}

@article{rudin2019stop,
  title={Stop explaining black box machine learning models for high stakes decisions and use interpretable models instead},
  author={Rudin, Cynthia},
  journal={Nature machine intelligence},
  volume={1},
  number={5},
  pages={206--215},
  year={2019},
  publisher={Nature Publishing Group UK London}
}

@inproceedings{he2016ups,
  title={Ups and downs: Modeling the visual evolution of fashion trends with one-class collaborative filtering},
  author={He, Ruining and McAuley, Julian},
  booktitle={proceedings of the 25th international conference on world wide web},
  pages={507--517},
  year={2016}
}

@article{wang2024making,
  title={Making Non-overlapping Matters: An Unsupervised Alignment enhanced Cross-Domain Cold-Start Recommendation},
  author={Wang, Zihan and Yang, Yonghui and Wu, Le and Hong, Richang and Wang, Meng},
  journal={IEEE Transactions on Knowledge and Data Engineering},
  year={2024},
  publisher={IEEE}
}

@inproceedings{cui2020herograph,
  title={HeroGRAPH: A Heterogeneous Graph Framework for Multi-Target Cross-Domain Recommendation.},
  author={Cui, Qiang and Wei, Tao and Zhang, Yafeng and Zhang, Qing},
  booktitle={ORSUM@ RecSys},
  year={2020}
}




\end{document}